\documentstyle[epsfig,12pt]{article}
\newcommand{\eqn}[1]{eq.(\ref{#1})}
\newcommand{\fig}[1]{Fig. \ref{#1}}
\newcommand{\nn}{\nonumber}
\newcommand{\bq}{\begin{equation}}
\newcommand{\eq}{\end{equation}}
\newcommand{\bqa}{\begin{eqnarray}}
\newcommand{\eqa}{\end{eqnarray}}
\def\tev{\begin{mathrm} TeV \end{mathrm}}
\def\mev{\begin{mathrm} MeV \end{mathrm}}
\def\gev{\begin{mathrm} GeV \end{mathrm}}
\def\fb{\begin{mathrm} fb \end{mathrm}}
\def\kg{\kappa_\gamma}

\def\lg{\lambda_\gamma}
\def\ptg{p_T^\gamma}
\def\ptj{p_T^{jet}}
\def\L{{\cal L}}
\def\to{{\rightarrow}}
\begin{document}
\begin{titlepage}
\begin{flushright}
CERN-TH.6753/92\\
DTP/92/92
\end{flushright}
\vskip 2cm
\begin{center}
{\large \bf { $W + \gamma + jet$ production as a test of
              the electromagnetic couplings of $W$ at LHC and SSC } }
\vskip 1cm

{\sl
 F.K. Diakonos$^a$, O. Korakianitis$^{a,b}$, C.G. Papadopoulos$^c$ \\
      C. Philippides $^{a,d}$, W.J. Stirling$^e$ }
\end{center}
\vskip 1cm
\begin{abstract}
The reaction  $pp \to W \ +\ \gamma\ +\ jet \ +\ X$
is considered at centre-of-mass
energies $\sqrt{s} = 16$ and $40\;\tev$, including
anomalous three-gauge-boson couplings
$\kappa$ and $\lambda$. The possibility of obtaining limits on these
quantities by comparison with the standard model is
investigated. The radiation zero
properties of the subprocess matrix elements are studied.
\end{abstract}
\vspace{1cm}
CERN-TH.6753/92\\
December 1992\\
\hrule\smallskip
\nopagebreak
{\footnotesize \noindent
a: Institute of Nuclear Physics, NRCPS \lq\lq Democritos",
GR-153 10 Athens, Greece. \\
b: Dept. of Physics, Royal Holloway and Bedford New College,
Univ. of London, Egham, Surrey TW20 OEX, UK.\\
c: CERN Theory Division, CH-1211 Geneva 23, Switzerland.\\
d: Department of Physics, New York University, NY 10003, USA. \\
e: Depts of Mathematical Sciences and
Physics, Durham University, South Road, Durham DH1 3LE,
UK.}
\end{titlepage}
\newpage
\par
One of the main features of the non-abelian character of
the $SU(2)_L \otimes
U(1)_Y$ standard model (SM), is the existence of trilinear and
quadrilinear gauge boson couplings with definite properties.
The precise form of these vertices, as specified in the SM,
has not yet been
verified experimentally. It is therefore important to measure
these couplings
and look for possible deviations from the SM, which would
provide hints for new physics such as compositeness of the
intermediate vector bosons.
\par
The most general C,P,T and $U(1)$ gauge invariant effective Lagrangian
describing three- and four-gauge-boson interactions can be written
as follows:
\begin{equation}
\L_{int} = \L^{(3)}_{SM} + \L^{(4)}_{SM} + \L^{(3)}_{\kappa} +
\L_{\lambda}\label{lagr}
\end{equation}
where the first two terms are the familiar SM three- and four-gauge-boson
interactions and their deviations are described by the last two
\begin{eqnarray}
\L^{(3)}_{\kappa} &=& -ie(\kappa -1)
W_{\mu}^{\dagger}W_{\nu} F^{\mu \nu} \\
\L_{\lambda} &=& -ie \frac{\lambda}{M_W^2} G_{ \mu \nu}
G^{\dagger \nu \rho} F^{\mu}_{\rho}
\end{eqnarray}
where
\begin{eqnarray}
F_{\mu \nu} &=& \partial_{\mu} A_{\nu} - \partial_{\nu}A_{\mu} \\
W_{\mu \nu} &=& \partial_{\mu} W_{\nu} - \partial_{\nu}W_{\mu} \\
G_{\mu \nu} &=& W_{\mu \nu} -ie (A_{\mu} W_{\nu} - A_{\nu} W_{\mu})
\end{eqnarray}
with $W_{\mu}$ the $W$ field and $A_{\mu}$ the photon field.
The parameters
$\kappa$ and $\lambda$, which enter both in the three- and
four-gauge-boson
couplings, are related to the anomalous magnetic dipole and
electric quadropole moments of $W$ by
\begin{equation}
\mu = \frac{e}{2M_W} (1+\kappa + \lambda) ,\ Q= \frac{e}{M^2_W} (\lambda
- \kappa)
\end{equation}
The SM has at {\it tree order}: $\kappa =1$ and $\lambda =0$.
The three- and four-gauge-boson vertices $W^+W^- \gamma$ and
$W^+W^- \gamma\gamma$,
respectively, implied by the Lagrangian of \eqn{lagr} are given
in ref. \cite{wgg}.
\par
The phenomenology of anomalous couplings in $W$ production at future $pp$
colliders
\begin{eqnarray}
 LHC:  \ \sqrt{s} &=& 16\;\tev , \ {\cal L}  =  100\; \fb^{-1} /year , \\
 SSC:  \ \sqrt{s} &=& 40\;\tev , \ {\cal L}  =  10\; \fb^{-1} /year .
\end{eqnarray}
has already been studied extensively in refs. \cite{wgg,zep,cort}.
The main purpose has been
to limit the values of the parameters $\kappa$ and
$\lambda$, which describe the deviation from the SM predictions.
Since these
parameters also receive contributions from higher-order corrections
to the SM,
it is of great importance to approach phenomenologically the limit
set by higher-order corrections. The large number of
events expected in the
strong process $pp \rightarrow W^{\pm} \ +\ \gamma\ +\ jet  \ +\ X$
offers large
statistics and therefore gives a better perspective to obtain more
stringent
bounds on the parameters $\kg,\ \lg$ than the previously examined process
$pp \rightarrow W^{\pm} +\ \gamma\ + \gamma \ +\ X$ \cite{wgg},
through the three-gauge-boson vertex entering
the Feynman diagrams (see \fig{diag}).
\par
{}From the three-gauge-boson vertex given in eq.(3) of ref.
\cite{sw},  we see that $\lambda$ is
multiplied by factors of the form ${p_{\pm}^2}/{M_W^2}$,
which are equal to 1 for on shell $W$'s but large for
processes involving virtual $W$'s. This
motivates us to consider $W \gamma\ jet$ production at $pp$
 colliders,
since in this process off-shell W's enter into several Feynman
diagrams and
thus we expect this reaction to be very sensitive to the parameter
$\lambda$.
This is actually the case and as we will show below, this reaction
provides
us with more stringent limits on $\lambda$ than $\kappa$, of the order
$5 \times 10^{-2}$, which is of the order of the 1-loop SM values
\cite{1loo}.
\par
Specifically, we consider the process:
$pp \rightarrow W^{\pm} \gamma \ jet+X$.
The cross section in the parton model is given by the usual
convolution of a subprocess cross section with parton distributions:
\begin{eqnarray}
{\sigma}(pp \rightarrow W^{\pm} \gamma\ jet\ +X)  &=& \sum_{a,b,c=q,g}
\int_0^{1}dx_a dx_b f_{a/p}(x_a,Q^2)f_{b/p}(x_b,Q^2)  \nonumber \\
&\times&
\hat{\sigma}(a(p_1)b(p_2) \rightarrow W(p_5) \gamma(p_3)c(p_4))
\end{eqnarray}
For the parton distributions we use the KMRS(B$_0$) set \cite{kmrs},
with $\Lambda_{\overline{\rm MS}}^{(4)} = 190\ \mev $ and
$Q^2= \hat{s} = x_a x_b s$.
For the electroweak parameters we use
$\sin^2{\theta}_W=0.23$, $M_W=80\ \gev$ and $\alpha_{em}^{-1}= 128 $.
\par
The Feynman diagrams contributing to the subprocess $u \ g \rightarrow
W^+ \gamma \ d$ are shown in \fig{diag}. The remaining five subprocesses
contributing to the total cross section, namely
\begin{eqnarray}
d g \to W^- \gamma u \nonumber \\
\bar{u} g \to W^- \gamma \bar{d} \nonumber \\
\bar{d} g \to W^+ \gamma \bar{u}  \nonumber \\
u \bar{d} \to W^+ \gamma g \nonumber \\
\bar{u}d \to W^- \gamma g \: ,
\end{eqnarray}
are obtained by crossing.
The subprocess matrix elements are
evaluated using the E-vector product technique
\cite{sw} and the explicit expressions for the
helicity amplitudes will be given elsewhere \cite{rev}.
The usual tests of
gauge invariance (with respect to the photon as well as the gluon polarization
vector) and unitarity have succesfully been performed.
\par
In order to identify the final state we have to impose appropriate cuts
on the produced photon and jet for two reasons:
\begin{itemize}
\item[(a)] To avoid infrared and collinear singularities and
\item[(b)] To extract reliably the photon signal, i.e.,
to reduce significantly
the $\pi^0 \to \gamma \gamma$ background \cite{exp}.
\end{itemize}
We therefore use the following cuts
\begin{eqnarray}
|\eta_{\gamma}| &\leq& 1.5 \: , \: \mid \eta_{jet} \mid \leq 2.5, \nn\\
\ptg &\geq & 50\ \gev \: , \; \ptj \geq 20\ \gev \nonumber \\
\Delta R_{\gamma -jet} &=& \sqrt{(\phi_\gamma-\phi_j)^2
+(\eta_\gamma - \eta_j)^2}
\geq 0.4 \: .
\label{cuts}
\end{eqnarray}
We also include a factor of $2/9$ for the leptonic branching ratio
for the decay of the W, in order to avoid the large QCD background.
Since at
this moment we are interested in examining the sensitivity of the
process to
$\kappa$ and
$\lambda$, rather than extracting precise bounds on these parameters,
we do
not address several questions concerning the experimental efficiency
for the
identification of the $W^{\pm} \gamma \ jet$ final state. We expect such
uncertainties to factor out when we take
ratios of differential to total cross sections, as in
\fig{dif}.
\par
We first consider the total cross section, subject to the cuts of
\eqn{cuts}
as a function of the anomalous couplings, $\sigma( \kappa,
\lambda )$. The cross section is a polynomial of degree two in
$\kappa-1$ and $\lambda$,
\bq
\sigma(\kappa,\lambda)=\sigma_0+\sigma_1^\kappa(\kappa-1)+
\sigma_2^\kappa(\kappa-1)^2+\sigma_1^\lambda\lambda+
\sigma_2^\lambda\lambda^2+\sigma^{\kappa\lambda}(\kappa-1)\lambda
\eq
In Tables 1 and 2, we give the coefficients of the total cross sections
corresponding to the above expansion.
\begin{table}[th]
\begin{center}
\begin{tabular}{|c|c|c|c|c|c|} \hline
$\sigma_0$ & $\sigma_1^\kappa$ & $\sigma_2^\kappa$ & $\sigma_1^\lambda$
& $\sigma_2^\lambda$ & $ \sigma^{\kappa \lambda}$ \\ \hline
2.1 & 0.42 & 0.87 & 0.31 & 0.55 & 0.29 \\ \hline
\end{tabular}
\caption[.] {Cross-section coefficients in picobarns, for
$\sqrt{s}=16$~TeV. }
\end{center}
\end{table}

\begin{table}[th]
\begin{center}
\begin{tabular}{|c|c|c|c|c|c|} \hline
$\sigma_0$ & $\sigma_1^\kappa$ & $\sigma_2^\kappa$ & $\sigma_1^\lambda$
& $\sigma_2^\lambda$ & $ \sigma^{\kappa \lambda}$ \\ \hline
6.5 & 1.3 & 2.3 & 0.54 & 2.8 & 1.0 \\ \hline
\end{tabular}
\caption[.]{Cross-section coefficients in picobarns, for
$\sqrt{s} = 40$~TeV. }
\end{center}
\end{table}

In \fig{cont} we show contours of equal cross sections in the
$\kappa-\lambda$ plane, corresponding to two standard deviations from
the SM values. The bounds obtained on each of the anomalous
couplings, are given for LHC by:
\bqa
0.5  &\le  \kappa \le ~0.54~~,~~~0.96  &\le  \kappa \le 1.02 \nn \\
-0.6 &\le  \lambda \le -0.52~~,~~-0.04 &\le  \lambda \le 0.04
\eqa
and for SSC:
\bqa
 0.4  &\le\kappa \le 0.48~~,~~0.96 &\le \kappa \le 1.04 \nn \\
-0.2 &&\le\lambda \le 0.05
\eqa
\par
In order to obtain bounds on the anomalous couplings from the
total event rate we have to take into account both experimental
(luminosity, acceptance, ...)
and theoretical (higher-order corrections, parton distributions, ...)
uncertainties on the measured and predicted cross sections.
{}From this point of view, it is
more reliable to study deviations from the expected {\it
differential} cross sections, where the overall normalization can be
factored out. Several quantities are likely to be useful in this
respect, but the one that is probably the simplest to measure is the
distribution of the {\it photon transverse momentum}, $\ptg$.
Since this is
the momentum entering the three-gauge-boson vertex and it
is multiplied by the
anomalous couplings $\kappa,\ \lambda$, we expect this distribution
to show
the bigest deviation from the SM result in magnitude as well as shape,
especially at higher values of the photon transverse momentum and
therefore of
$\sqrt{\hat{s}}$, where violation of unitarity becomes more apparent.
In \fig{dif}, we show the normalized distributions
$\frac{1}{\sigma} \frac{d\sigma}{d\ptg}$ and
$\frac{1}{\sigma} \frac{d\sigma}{d\ptj}$
at $\sqrt{s}=16~\tev$
and $\sqrt{s}=40~\tev$, respectively,
for $(\kappa,\lambda) = (1,0), (1.5,0)$ and $(1,0.5)$.
We see that the jet distributions are less sensitive to
the anomalous couplings. This suggests that the
photon transverse momentum is the main tool to analyse the
electromagnetic couplings of the $W$.
\par In order to compare with the previous calculations on $W\gamma
\gamma$ production, we show in \fig{comp} the ratio $\frac{\sigma(
\kappa,\lambda)}{\sigma_0}$ as a function of $\Delta\kappa=\kappa-1$ and
$\lambda$, where $\sigma_0=\sigma(\kappa=1,\lambda=0)$. We see that the
sensitivity of the process $pp\to W\gamma\gamma$ is much larger than
that of the process $pp\to W\gamma jet$. This is due to the fact that
the cross section for the former process is a quartic polynomial in
$\Delta\kappa$ and $\lambda$ whereas for the latter we have at most
quadratic dependence. Nevertheless, the process under consideration
gives a cross section which is orders of magnitude larger that the
$W\gamma\gamma$ one and thus offers a large statistics channel for the
study of anomalous $W$ couplings.
\par
A final point that we wish to study is the impact of the
presence of a {\it radiation zero} in the $q_i \bar{q_j} \to W\gamma g$ matrix
element.
It is well known \cite{brod}
that the leading order $u \bar d\to W^+ \gamma$ amplitude
vanishes when the centre-of-mass scattering angle of the photon
satisfies $\cos\theta_\gamma = 1+ 2 Q_d = 1/3$. Less well known
\cite{brod}
is the fact that this zero is unchanged when additional neutral
particles are emitted {\it in the same direction as the photon}.
In both cases, the zero disappears when additional anomalous three-
and four-gauge-boson couplings are included in the Lagrangian.
In the present context this means that the amplitudes such as
$u \bar d \to W^+ \gamma g$ vanish when (a) the photon and the gluon
are collinear ($M_{\gamma g} = 0$), (b) $\cos\theta_{\gamma g} = 1/3$
in the
$u\bar d$ centre-of-mass, and (c) $\kappa = 1,\ \lambda = 0$.
There are two issues here. First, the existence of a zero under these
conditions is a powerful check on the matrix element calculation and,
secondly,
can the zero be observed, and if so can it provide more stringent
limits on the anomalous couplings than the total cross section and
$p_T$ distribution measurements described above?
\fig{zero}
shows the matrix element squared $|M(u \bar d\to W^+\gamma g)|^2$
for $\sqrt{\hat s} = 1~\tev$ and equal photon and gluon momenta
in the final state ($p_3^\mu = p_4^\mu$), as a function of the
common photon-gluon centre-of-mass scattering angle $\theta_{\gamma g}$
for different choices of the anomalous couplings $\kappa$ and $\lambda$,
$(\kappa,\lambda)=(1,0),(1.1,0)(1,0.1),(1,1)$.
The zero in the Standard Model case is immediately apparent, and the
anomalous couplings are seen to fill in the dip caused by the
radiation zero.
In the region of the zero there is evidently a high-sensitivity to these
couplings.
\par
Unfortunately it is in practice very difficult to identify the
$W\gamma$ radiation zero in high-energy hadron-hadron collisions
\cite{cort,noze}.
The main problems are that (a) the zero occurs in the subprocess
centre-of-mass frame which is different event by event from the
laboratory frame, (b) the centre-of-mass frame is difficult
to construct from the final state momenta because there is
an undetected neutrino in the leptonic decay of the $W$, and (c) at
high energy proton-proton colliders, the $u \bar d$ process
is overwhelmed by processes involving an initial state gluon,
e.g. $u g \to W^+ \gamma d$. (Explicit computation shows that
$qg$ scattering processes represent $88\%$ ($92\%$) of the total
$W+\gamma+jet$ cross section at LHC (SSC) energies.)
Not only do these latter processes have
no radiation zero but  they are also singular when the final state
photon and jet are collinear! These singularities are only
regulated when  a proper next-to-leading order definition of  a jet
is implemented (for example, as energy inside a cone).
In other words, it seems impossible  either theoretically or
experimentally to identify a parallel photon and gluon in the final
state.

\par
In conclusion, we have shown that the study of the
$pp \rightarrow W^+ \gamma \ jet + \ X$ reaction will provide at future
high energy proton-proton colliders many events
to test the electroweak three- and four-gauge-boson couplings of the
standard
model, which are a direct consequence of the $SU(2)_L \otimes U(1)_Y$
gauge symmetry. Furthermore, the radiation zero of the amplitudes is still
there for a particular phase space configuration and is very sensitive to
the anomalous couplings $\kappa,\ \lambda$. In an ideal experimental
situation where the subprocesses $u\bar{d} \to W^+ \gamma\ g$ and
$d\bar{u} \to W^- \gamma\ g$ could be isolated and their cross sections
measured
in the partonic centre-of-mass frame, this zero would provide us with
arbitrarily high sensitivity to measure the anomalous couplings.

\bigskip
\bigskip
\noindent{\large\bf Acknowledgements}

\medskip
\noindent This work was sponsored in part by the British Council
in Athens. WJS is grateful to the Institute of Nuclear Physics
``Democritos" for hospitality during the course of this work.

\end{document}